\documentclass[aps,pre,twocolumn,showpacs]{revtex4} 
\usepackage{graphicx}
\usepackage{color}
\usepackage{hyperref}
\usepackage[justification=centering]{caption}
\usepackage[sans]{dsfont}
\usepackage{subfig}

\usepackage{hyperref}
\usepackage{dsfont}
\usepackage{amsmath}
\usepackage{amssymb}
\usepackage{lipsum}
\begin{document}
\title{Performance Comparison of Quantum Otto and Stirling Engines Using Atom-Photon Interactions}
\author{Indrajith.V.S}
\email{indraphysics08@gmail.com}
\affiliation{Qdit Labs Pvt. Ltd, Bangalore - 560092, Karnataka India.}
\bigskip
\begin{abstract}
This article presents a comparative analysis of quantum Otto and Stirling engines using atom-photon interactions as the working substance. Two models are considered: a two-level Jaynes-Cummings system and a four-level atomic system confined in an optical cavity. The thermodynamic cycles are analyzed, highlighting the role of critical points, quantum correlations, and atom-field interactions in determining efficiency and work output. Our findings provide insights into the advantages and limitations of QHEs, offering a deeper understanding of energy conversion in quantum systems.
\end{abstract}
\maketitle
\section{Introduction}
Throughout history, humanity's progress has been driven by remarkable innovations, with fire and the wheel standing as early milestones. Among these, the development of machines—particularly thermal machines—has played a pivotal role in shaping civilization. The ability to harness and convert heat energy into mechanical work marked the dawn of industrialization, revolutionizing transportation, manufacturing, and energy production. Over centuries, while the materials and mechanisms of these machines have evolved, their fundamental purpose has remained unchanged: the efficient transformation of thermal energy into useful work.

Quantum heat engines (QHEs) operate based on quantum mechanical principles, distinguishing them from classical thermal machines. Unlike classical engines, which rely on macroscopic thermodynamic processes, QHEs incorporate uniquely quantum effects such as superposition, coherence, and entanglement, fundamentally altering how energy is transferred and utilized. Since the introduction of the first theoretical model in 1959 \citep{qhe1}, extensive research has led to the development of numerous QHE models, each leveraging different quantum properties to enhance performance.

Studies have highlighted several advantages of quantum thermal machines over their classical counterparts \cite{qhe2, qhe_book, qhe3}. One of the key factors contributing to this advantage is the role of quantum coherence and entanglement in thermodynamic processes. These quantum correlations can enhance the efficiency and power output of heat engines beyond what is achievable with classical systems \cite{adv1, adv2, adv3, adv4, adv5, adv6, adv7, adv8}. Remarkably, research has shown that the efficiency of QHEs can even surpass the classical Carnot limit without violating the second law of thermodynamics  \cite{coh_bath1, coh_bath2, coh_bath3, coh_bath5, sq_bath1, sq_bath2, sq_bath3, non_mar_bath1, corr_bath1}. This breakthrough is possible due to the exploitation of quantum resources, which modify the traditional constraints imposed by classical thermodynamics.

Researchers have developed various quantum heat engine (QHE) models leveraging distinct quantum properties. Coherent thermal baths enhance energy transfer via quantum coherence \cite{coh_bath1, coh_bath2, coh_bath3, coh_bath5}, while squeezed thermal baths utilize non-equilibrium states to reduce entropy and boost work extraction \cite{sq_bath1, sq_bath2, sq_bath3}. Non-Markovian dynamics introduce memory effects that sustain coherence and reduce dissipation \cite{non_mar_bath1}. Quantum-correlated baths exploit entanglement between thermal reservoirs, enabling additional energy extraction beyond classical limits \cite{corr_bath1}.


In this article, we present a comparative study of quantum Otto and Stirling engines, employing an atom-photon interaction as the working medium. Specifically, we consider two models: the two-level Jaynes-Cummings system and a four-level atomic system confined within an optical cavity. The interaction of atoms with photons, which facilitates energy exchange and influences thermodynamic cycles, forms the core of our investigation. We systematically analyze the operation of these engines within the Otto and Stirling cycles and compare their performance under different physical conditions. By examining the thermodynamic characteristics of these two systems, we aim to gain deeper insights into the advantages and limitations of quantum thermal machines in harnessing atom-light interactions for energy conversion.
\section{Operation cycles of Quantum heat engine}
\subsection{Working of Stirling engine}
A Stirling engine operates on a closed-loop thermodynamic cycle known as the Stirling cycle \citep{Stirling1, Stirling2, Stirling3}. Unlike many other heat engines that rely on continuous combustion processes, the Stirling engine functions by cyclically compressing and expanding a fixed quantity of working substance within a sealed chamber. In contrast, a quantum Stirling engine leverages quantum mechanics by manipulating the quantum states of its working substance, typically atoms or molecules, to optimize energy conversion.

The operation of a quantum Stirling engine involves four distinct stages: two isothermal strokes and two isochoric strokes. During the isothermal stroke, the control parameter changes while the temperature remains constant. In contrast, during the isochoric stroke, the control parameter stays consistent. This four-stroke thermal machine can be characterized by the following stages:

\textbf{Stage A $\rightarrow $ B:}

This constitutes an isothermal stroke, where the temperature is meticulously maintained by coupling the system to a heat bath, allowing the control parameter to vary quasi-statically. It is crucial to note that the system sustains thermal equilibrium with the heat bath throughout this stroke. Simultaneously, the system yields a certain amount of work. The heat exchanged with the hot reservoir is computed as:
\begin{equation}
Q_{AB} = T_h (S_B - S_A),\label{St_JC_ab}
\end{equation}
where $S_k$ represents the von Neumann entropy of the system.

\textbf{Stage B $\rightarrow $ C :}

This encompasses an isochoric stroke, characterized by a constant control parameter. In this phase, the system disengages from the hot reservoir and establishes a connection with a cold reservoir. Given the unaltered energy levels, no mechanical work is executed, and the heat exchange is articulated as:
\begin{equation}
Q_{BC} = \sum_i E_{Bi} (p_{Ci} - p_{Bi}).\label{St_JC_bc}
\end{equation}
\textbf{Stage C $\rightarrow$ D:}

In this isothermal compression phase, the system stays in contact with the cold bath, and the control parameter is tuned to revert to the initial state. The exchanged heat is defined as:
\begin{equation}
Q_{CD} = T_c (S_D - S_C).\label{St_JC_cd}
\end{equation}
\textbf{Stage D $\rightarrow$ A:}

In this concluding stage of isochoric compression, the system transitions from the cold reservoir to being coupled with the hot reservoir, maintaining a constant control parameter at the lower level. The heat exchange is expressed as:
\begin{equation}
Q_{DA} = \sum_i E_{Ai} (p_{Ai} - p_{Di}).\label{St_JC_da}
\end{equation}
The total heat absorbed from the hot reservoir is the sum of heat exchanged in stages I and IV, given by $ Q_h = Q_{AB} + Q_{DA}$, while $ Q_{c} = Q_{BC} + Q_{CD}$ represents the total heat exchanged with the cold reservoir.
\subsection{Modelling quantum Otto engine}
The Otto engine functions based on the Otto cycle, a thermodynamic cycle that outlines the process by which the engine transforms energy stored in the working substance into mechanical work. This fundamental operation adheres to the principles of the Otto cycle, encompassing stages such as compression, heating, expansion, and cooling.

The quantum Otto engine is characterized by two adiabatic strokes and two isochoric strokes, detailed as follows:\\

\textbf{Stage $a \rightarrow b$:} 

The initial stage involves an adiabatic process where the system is prepared at the initial temperature $T_h$, varying the control parameter quasistatically. This raises the energy levels ($E_a \rightarrow E_b$) while maintaining a constant probability distribution ($p_a$), resulting in net work done on the system.

\textbf{Stage $b \rightarrow c$:} 

Subsequently, an isochoric stage follows, conserving energy levels by keeping the control parameter constant. The system is in contact with the heat source, absorbing heat and causing a change in the probability distribution ($p_a \rightarrow p_b$). The heat exchange is given by:
\begin{equation}
Q_h = U_c - U_b = \sum_i E_{bi} (p_{bi} - p_{ai}).
\end{equation}
where $U$ represents the internal energy of the system. 

\textbf{Stage $c \rightarrow d$:}

In the subsequent adiabatic compression, the control parameter is decreased back to its initial value. This leads to a decrease in energy levels ($E_b \rightarrow E_a$)with the probability distributions remaining constant, thus resulting in no work done.

\textbf{Stage $d \rightarrow a$:}

 Finally, in the isochoric compression stage, the system is coupled to a cold bath, and the probability distribution returns to the initial state, resulting in heat exchange which is given by:
 \begin{equation}
 Q_c = U_a - U_d = \sum_i E_{ai} (p_{ai} - p_{bi}).
 \end{equation}
 The total heat exchanged equates to the work done, given by $W= Q_c + Q_h$.
\section{Two level quantum heat engine using Jaynes Cummings Model}
\subsection{The Model}
To start, we conceptualize a quantum heat engine utilizing a two-level system. Figure (\ref{JC}) presents a schematic representation of this two-level system, characterized by its ground state $\lvert g \rangle$
and excited state $\lvert e \rangle$. A particle confined within a cavity can transition between these energy states, facilitated by adjusting the frequency of an external field. Such adjustments induce oscillations of the particle between these two states. A prominent example of a two-level system arising from light-matter interaction is the Jaynes-Cummings model \citep{JC_Model}.

The Jaynes-Cummings model stands as a cornerstone in quantum optics, delineating the interplay between a two-level quantum system (typically depicted as an atom) and a quantized electromagnetic field mode (often symbolized by a single-mode cavity). This model furnishes an essential framework for elucidating the dynamics of quantum systems in tandem with quantized radiation fields.
\begin{figure}[!ht]
\centering
 \includegraphics[width=0.8\linewidth]{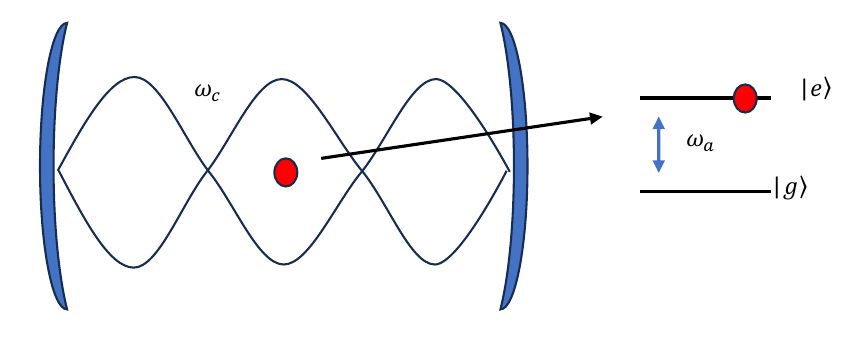}
 \caption{Jaynes Cummings Model}
\label{JC} 
\end{figure}

The Hamiltonian of the Jaynes-Cummings model is expressed as \citep{JC_Model}:
\begin{equation}
H_{J} = \hbar \omega_c a^\dagger a + \hbar \omega_a \frac{\sigma_z}{2} + \hbar g (a \sigma_+ + a^\dagger \sigma_-)\label{JC hamil}
\end{equation}
in the basis $ \lvert n -1,\, 0\rangle $, $ \lvert n,\,1\rangle $, where $\lvert n \rangle $ describes the state of $n$ photons. Here, $a$ and $a^\dagger$ correspond to creation and annihilation operators, and $\sigma_+$ and $\sigma_-$ are ladder operators. The energy eigenvalues corresponding to the Jaynes-Cummings model are given by \citep{JC_Model}:
\begin{equation}
E_{1,2} = (n+\frac{1}{2}) \omega_c \pm \hbar \sqrt{\Delta^2 + (2g)^2 (n+1) }
\end{equation}
where $\Delta = \omega_a - \omega_c$, and $\omega_{a,c}$ are the atomic transition frequency and cavity mode frequency, respectively. In this context, the working substance is the atom trapped in the cavity, coupled with a heat source. The term $g$ denotes the coupling between the atom and the field and serves as the chosen control parameter.

\subsubsection*{Stirling Engine}
The system begins in state $A$ in thermal equilibrium with a heat bath at temperature $T_h$ and an atom-field coupling constant $g_1$. As the system undergoes an isothermal process, transitioning to a coupling constant $g_2$, both work and heat absorption occur. The heat exchanged during this phase can be determined as:
\begin{equation}
Q_{AB} = [U_B(T_h,g_2) - U_A(T_h,g_1) ] T_h \log\left(\frac{Z_B}{Z_A}\right)
\end{equation}
where $U = \sum_i p_i E_i$ is the internal energy, and $Z = \sum_i p_i$ is the partition function. In the following isochoric process, the system is disconnected from the hot bath and connected to a cold reservoir at temperature $T_l$, leading to heat exchange as defined by equation (\ref{St_JC_bc}). Subsequently, the system undergoes an isothermal expansion stroke, where the coupling constant returns to its initial value, resulting in both work and heat exchange, calculated as:
\begin{equation}
Q_{CD} = [U_D(T_l,g_1) - U_A(T_l,g_2) ] T_l \log\left(\frac{Z_D}{Z_C}\right).
\end{equation}
The final stroke enables the system to re-engage with the hot bath, returning it to its initial condition and completing the cyclic process. The total heat exchanged with both the hot and cold baths is calculated as:
\begin{multline}
    Q_h = (U_B - U_D) + T_h \log\left(\frac{Z_B}{Z_A}\right), Q_c =\\ 
-(U_B - U_D) + T_c \log\left(\frac{Z_D}{Z_C}\right).
\end{multline}
This gives the total work done as $W = Q_h + Q_c =  T_h \log\left(\frac{Z_B}{Z_A}\right) +T_c \log\left(\frac{Z_D}{Z_C}\right)$. In the following analysis, we examine the influence of various parameters on the extraction of work and efficiency in a Stirling engine.  For convenience in our analysis, we set $\hbar = 1$.
\begin{figure}[!ht]
 \includegraphics[width=0.45\linewidth]{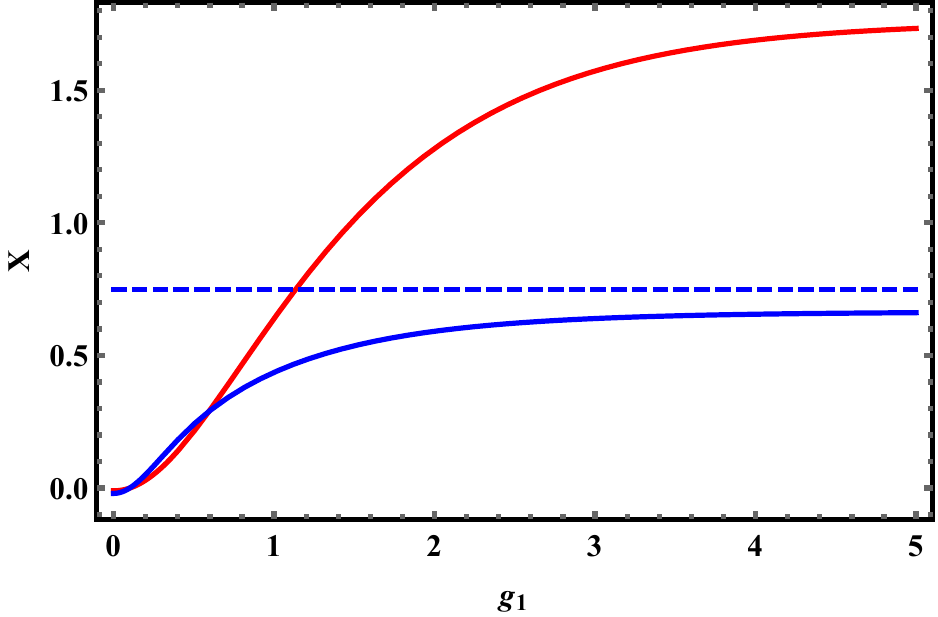}
 \includegraphics[width=0.45\linewidth]{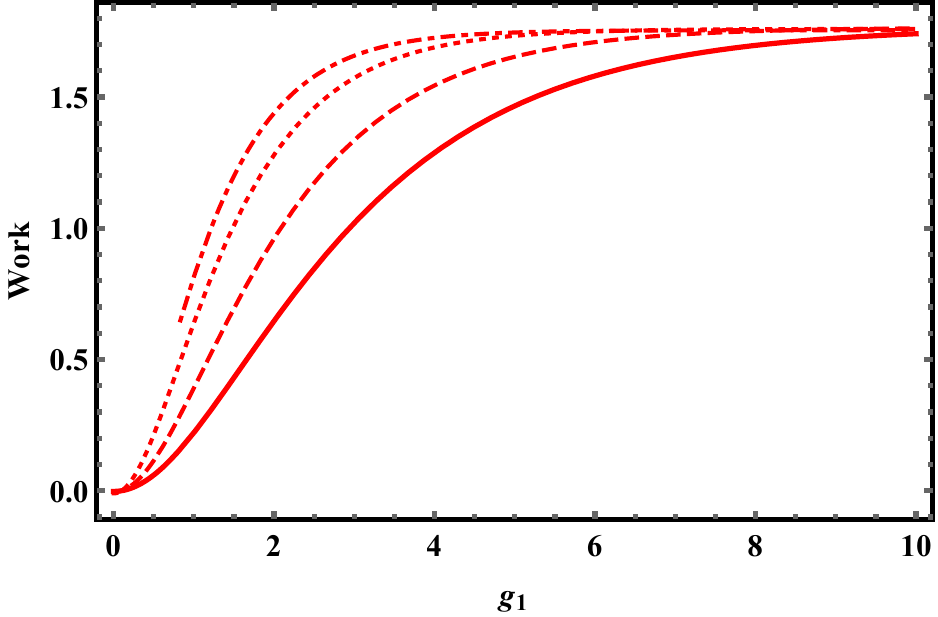}
 \caption{(i)Work (Red) and Efficiency (Blue) of Four level Stirling engine in comparison with the Carnot efficiency (Dotted) for $n=3, T_h = 4, T_c =1, g_2 =0.1,\omega_a = 3, \omega_c = 1$. (ii) Work as a function of atom-field interaction for different values of number of photons.}
\label{JC_work} 
\end{figure}

Figure [\ref{JC_work}] depicts the variation in work and efficiency with respect to the coupling constant. The trend indicates an increase in work with the coupling constant, eventually reaching a saturation point. The saturation value of work is unaffected by the number of photons within the cavity but does rely on the bath temperature. Conversely, a higher number of photons causes the work to achieve its maximum value even with a slight change in the coupling constant, as evident in Fig (ii).

The detuning frequency ($\omega_a - \omega_c$) significantly influences the maximum extractable work, with a lower detuning frequency resulting in higher work done. A parallel observation is made for efficiency, wherein it approaches the Carnot limit as the detuning frequency approaches zero.
\subsection{Otto engine}
The heat absorbed during the isochoric expansion is given by:
\begin{equation}
Q_h = U_c(T_l, g_l) - U_b(T_h, g_l).
\end{equation}
Here, the internal energy is defined as $U = \frac{2}{Z} e^{\frac{A}{T}}(A \cosh(\frac{k}{T})- k \sinh(\frac{k}{T}))$, where $A= (n+\frac{1}{2})\omega_c$ and $k= \frac{1}{2}\sqrt{\Delta^2 + (2g)^2 (n+1) }$.

The heat rejected to the heat bath is calculated as:
\begin{equation}
Q_c = U_a(T_h,g_h) - U_d(T_l,g_h).
\end{equation}
The work done is then determined as the sum of the total heat exchanged.
\begin{figure}[!ht]
\centering
 \includegraphics[width=0.7\linewidth]{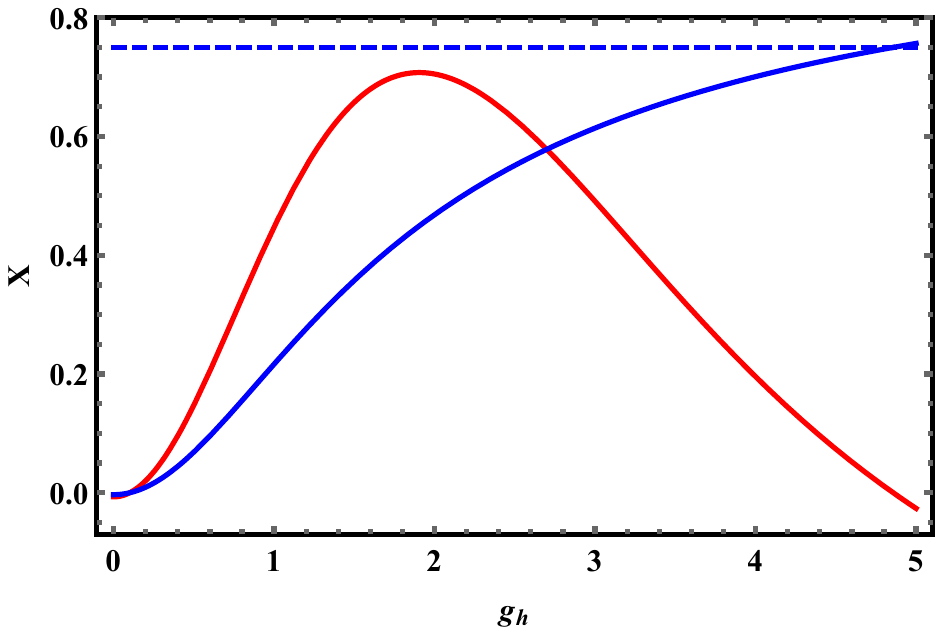}
 \caption{Work (Red) and Efficiency (Blue) of two level Otto engine in comparison with the Carnot efficiency (Dotted) for $n=3, T_h = 4, T_c =1, g_c =0.1,\omega_a = 3, \omega_c = 0.5$. }
\label{JC_Otto_work} 
\end{figure}

Figure (\ref{JC_Otto_work}) illustrates the variation of work and efficiency concerning the coupling parameter. Notably, positive work is achievable within a limited window of $g_h$ ,and this region of positive work is contingent on the lower coupling parameter $g_l$. The impact of other parameters mirrors that of the Stirling engine. A key distinction is that the work done saturates to a maximum value, and the efficiency reaches its peak below the Carnot efficiency. In contrast, the efficiency of the Otto engine can attain the maximum Carnot value when the work is positive. The peak value of the work is contingent on the temperature of the baths.

 A comparative analysis of the work done for both the Otto and Stirling engine is given below.
 \begin{figure}[!ht]
\centering
 \includegraphics[width=0.7\linewidth]{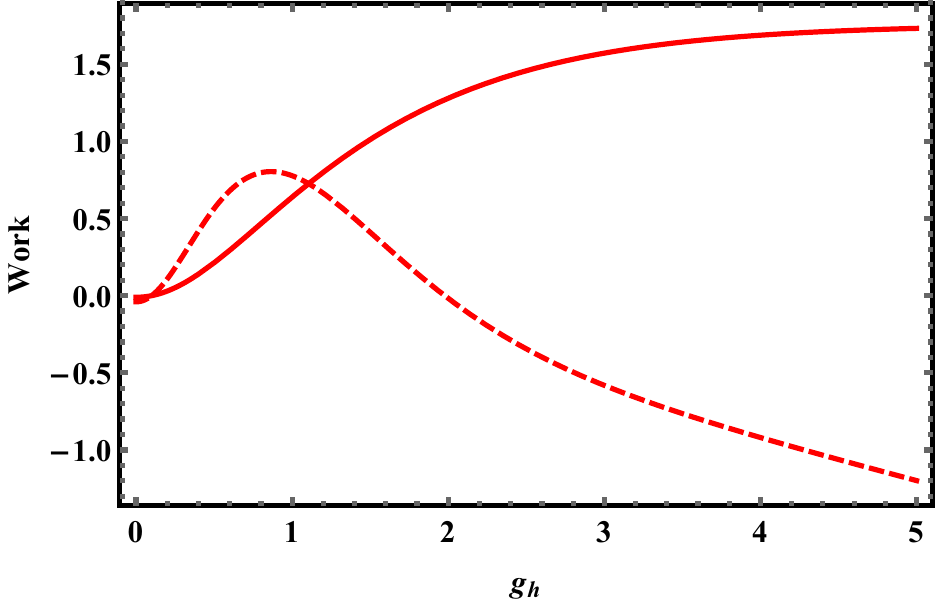}
 \caption{Work done in Otto engine (Dotted) and Stirling Engine (solid) and a function of atom-field coupling with $n=3, T_h = 4, T_c =1, g_c =0.1,\omega_a = 3, \omega_c = 0.5$. }
\label{JC_Comp} 
\end{figure}
\section{Four level system}
\begin{figure}[!ht]
\centering
 \includegraphics[width=0.8\linewidth]{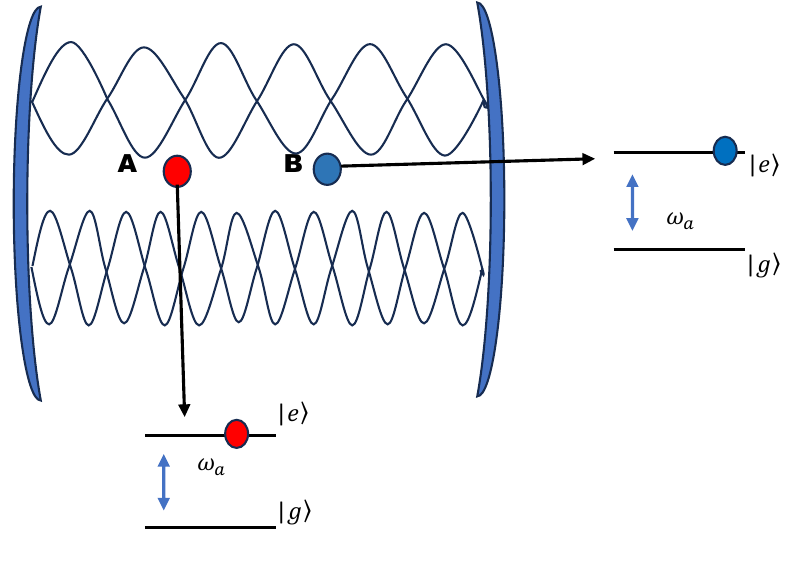}
 \caption{Tavis Cummings Model}
\label{JC} 
\end{figure}
\subsection{Stirling Engine}
In this study, we examine a system comprising two two-level atoms interacting with a single-mode cavity field, initially prepared in Fock states. The system's Hamiltonian, governing the atom-field interaction, is expressed as follows \citep{TC_model}:
\begin{equation}
H = \sum^2_{i =1} g(a \sigma_i^+ + a^\dagger \sigma_i^-) + 2 k(\sigma^+ \sigma^- + \sigma^- \sigma^+) + J \sigma_z \sigma_z \label{four level ham}
\end{equation}
Here $g$, $k $ and $J$ represent the atom-field coupling strength, dipole-dipole interaction strength, and Ising interaction strength, respectively. The corresponding eigenvalues of the Hamiltonian are given by  \citep{TC_model}:
\begin{eqnarray*}
E_1 = J && E_2 = -(J+2k) \\
E_3 = k + \alpha && E_4 = k - \alpha
\end{eqnarray*}
where $\alpha = \sqrt{(2n +1)2g^2 + (J-k)2}$. The partition function $\mathcal{Z}$ is defined as:
\begin{equation}
\mathcal{Z} = e^{(-J/T)}+e^{(J+2k/T)} + 2 e^{-k/T}\cosh(\alpha/T).
\end{equation}
Similarly, the internal energy $U$ is expressed as $U = J e^{-J/T} -(J+2k)e^{(J+2k)/T} +2 e^{-k/T}(k \cosh \frac{\alpha}{T} -\alpha \sinh \frac{\alpha}{T})$. 

The system described by the Hamiltonian in equation (\ref{four level ham}) serves as the working substance undergoing Stirling engine cycles, thereby generating work. The resulting work output and efficiency for the four-level system are depicted as follows:
\begin{figure}[!ht]
\centering
 \includegraphics[width=0.7\linewidth]{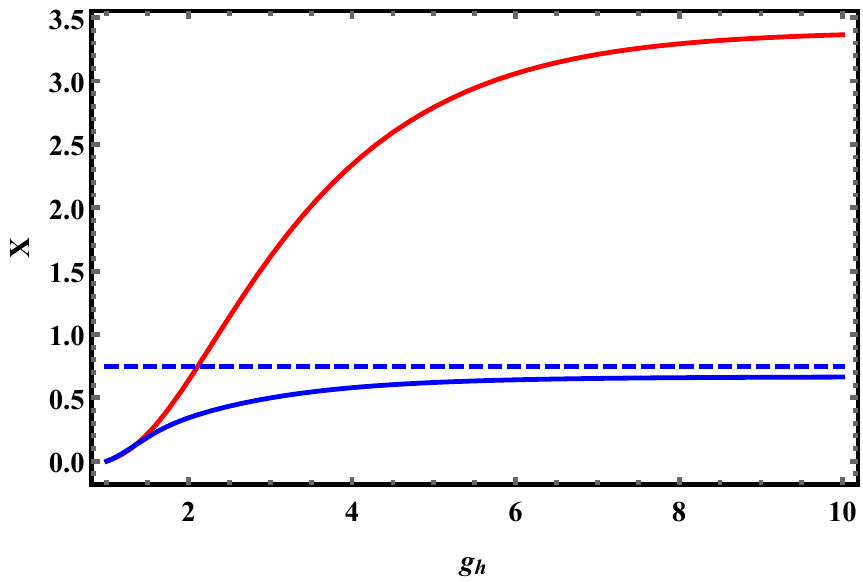}
 \caption{Work (Red) and Efficiency (Blue) of Four level Stirling engine in comparison with the Carnot efficiency (Dotted) for $n=1, T_h = 4, T_c =1, g_l =1, k = 1, J = 0.2$. }
\label{Four level_work} 
\end{figure}

The figure (\ref{Four level_work}) illustrates that, akin to the two-level system, the four-level heat engine exhibits a similar behavior. Both efficiency and work done increase with increasing atom-field coupling, reaching a maximum asymptotically. Remarkably, the number of photons within the cavity has no discernible effect on the work done. In contrast, the dipole-dipole interaction plays a pivotal role in determining the direction of work. Specifically, when the dipole interaction is less than 1 ($k \geq 1$), the work is consistently positive. The efficiency also follows a similar trend but is less influenced than the work done.
\subsection{Otto Engine}
The heat absorbed and rejected can be determined by assessing the change in internal energy during the isochoric expansion and compression. The internal energy for the four-stage Otto cycle is computed as follows:
\begin{equation}
U = \frac{1}{Z} \bigg(e^{-\beta J} + e^{\beta(J+2k)} + e^{-\beta k}[k \cosh(\beta \alpha) -\alpha \sinh(\beta \alpha)]\bigg).
\end{equation}

Here $\beta = \frac{1}{T}$. The subsequent discussion illuminates the impact of atom-field coupling on the work done and efficiency.
\begin{figure}[!ht]
\centering
 \includegraphics[width=0.7\linewidth]{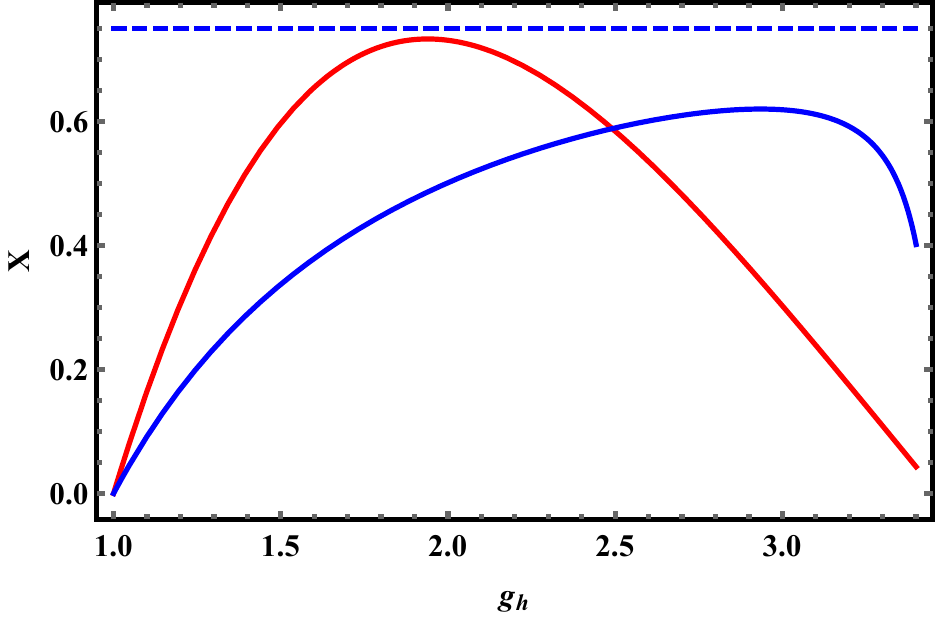}
 \caption{Work (Red) and Efficiency (Blue) of Four level Otto engine in comparison with the Carnot efficiency (Dotted) for $n=1, T_h = 4, T_c =1, g_l =1, k = 1, J = 0.2$. }
\label{Four level_otto} 
\end{figure}
In Figure (\ref{Four level_otto}), akin to the two-level quantum heat Otto engine, a finite region exists where work extraction from the system is feasible, beyond which the work becomes negative. The condition for positive work predominantly hinges on the atom-field coupling value, specifically requiring $g_l \geq g_h$. The maximum attainable work value is influenced solely by variations in the temperature of the baths. In contrast to the two-level system, the efficiency of the four-level system is notably lower and consistently falls short of reaching the maximum Carnot value.

\begin{figure}[!ht]
\centering
 \includegraphics[width=0.7\linewidth]{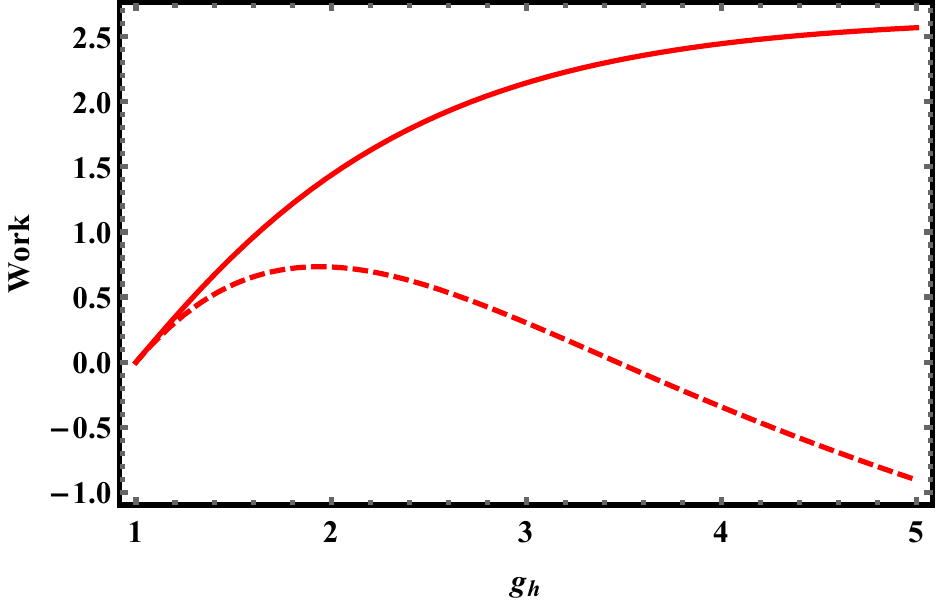}
 \caption{Work done in Otto engine (Dotted) and Stirling Engine (solid) and a function of atom-field coupling with $n=3, T_h = 4, T_c =1, g_l = 1, k = 0.1, J = 1$. }
\label{Four level_comp} 
\end{figure}

In Figure (\ref{Four level_comp}), a comparative analysis of the work performed by the Otto engine and the Stirling engine is presented. Both the two-level and four-level systems exhibit similar work behavior to that of the Otto engine, but with a smaller region of positive work compared to the Stirling engine. Additionally, the Stirling engine proves to be more efficient and capable of producing comparatively larger amounts of work.
\subsection{Role of entanglement}
The four level atom-field system defined in eq.(\ref{four level ham}) can be represented by the density operator as:
\begin{equation}
\rho = \sum_i p_i \lvert \psi_i \rangle \langle \psi_i \rvert
\end{equation}
where $\lvert \psi \rangle $ are the eigen vectors of the Hamiltonian which is given by
\begin{align*}
\lvert \psi_1 \rangle &= \frac{1}{\sqrt{2n+1}}\Big(-\sqrt{n}\lvert n+1,00\rangle + \sqrt{n+1} \lvert n-1, 11\rangle\Big)\\
\lvert \psi_2 \rangle &= \frac{1}{\sqrt{2}}\Big(-\lvert n,01\rangle +  \lvert n, 10\rangle\Big)\\
\lvert \psi_3 \rangle &= \frac{1}{2 \sqrt{\alpha(\alpha + (k-J))}}\Big(-2g\sqrt{n+1}\lvert n+1,00\rangle 
\\ &{\hspace{12pt}}+  2g\sqrt{n}\lvert n-1, 11\rangle + (\alpha +(k-J))[\lvert n,01\rangle +  \lvert n, 10\rangle]\Big)\\
\lvert \psi_4 \rangle &= \frac{1}{2 \sqrt{\alpha(\alpha - (k-J))}}\Big(-2g\sqrt{n+1}\lvert n+1,00\rangle 
\\ &{\hspace{12pt}}+  2g\sqrt{n}\lvert n-1, 11\rangle + (\alpha -(k-J))[\lvert n,01\rangle +  \lvert n, 10\rangle]\Big).
\end{align*}
The reduced density operator after tracing out the cavity field is given by:
\begin{equation}
\rho_{ab} = \frac{1}{\mathcal{Z}}
\begin{pmatrix}
\rho_{11} && 0 && 0 && 0\\
0 && \rho_{22} && \rho_{23} && 0\\
0 && \rho_{32} && \rho_{33} && 0\\
0 && 0 && 0 && \rho_{44}
\end{pmatrix} \label{themal state}
\end{equation}
where the elements are given by:
\begin{align*}
\rho_{11} &= \bigg(\frac{1}{2n+1}e^{-\beta J} + \frac{n+1}{2n+1}e^{-\beta k}[\cosh{(\beta \alpha)} \\
&+ \frac{(k-J)}{\alpha} \sinh(\beta \alpha)]\bigg)\\
\rho_{22} &= \rho_{33} = \frac{1}{2}\bigg(e^{\beta (J+2k)} + e^{-\beta k}[\cosh{(\beta \alpha)}\\
& - \frac{(k-J)}{\alpha} \sinh(\beta \alpha)]\bigg)\\
\rho_{23} &= \rho_{32} = -\frac{1}{2}\bigg(e^{\beta (J+2k)} + e^{-\beta k}[\cosh{(\beta \alpha)} \\
&- \frac{(k-J)}{\alpha} \sinh(\beta \alpha)]\bigg)\\
\rho_{44} &= \bigg(\frac{1}{2n+1}e^{-\beta J} + \frac{n}{2n+1}e^{-\beta k}[\cosh{(\beta \alpha)} \\
&+ \frac{(k-J)}{\alpha} \sinh(\beta \alpha)]\bigg).
\end{align*}
The entanglement of the above density operator in eq.(\ref{themal state}) is calculated using concurrence which is calculated as \citep{Cocurrence}
\begin{equation}
C(\rho) = 2 \,\text{max}\{\sqrt{\lambda_1} -\sqrt{\lambda_2} - \sqrt{\lambda_3} - \sqrt{\lambda_4}\}.
\end{equation}
Here $\lambda_i$ are the eigen values of $\rho \tilde{\rho}$, where $\tilde{\rho}$ is the spin flipped density operator. With this the concurrence of the thermal state in eq.{\ref{themal state}} is calculated as:
\begin{equation}
C(\rho_{ab}) = 2\, \text{max}\bigg\{0, \rho_{23}- \sqrt{\rho_{11}\rho_{44}}\,\bigg\}.
\end{equation}

In the following discussion, we examine the role of correlation (entanglement) in the various stages of a heat engine. It is evident that entanglement undergoes changes with temperature, wherein higher temperatures correspond to lower levels of correlation. It is noteworthy that entanglement significantly diminishes when the system is coupled to the hot bath. Here, we endeavor to analyze the fluctuation of correlation at each stage of the heat engine and explore its impact on the work output from the thermal machine.

\begin{figure}[!ht]
 \includegraphics[width=0.45\linewidth]{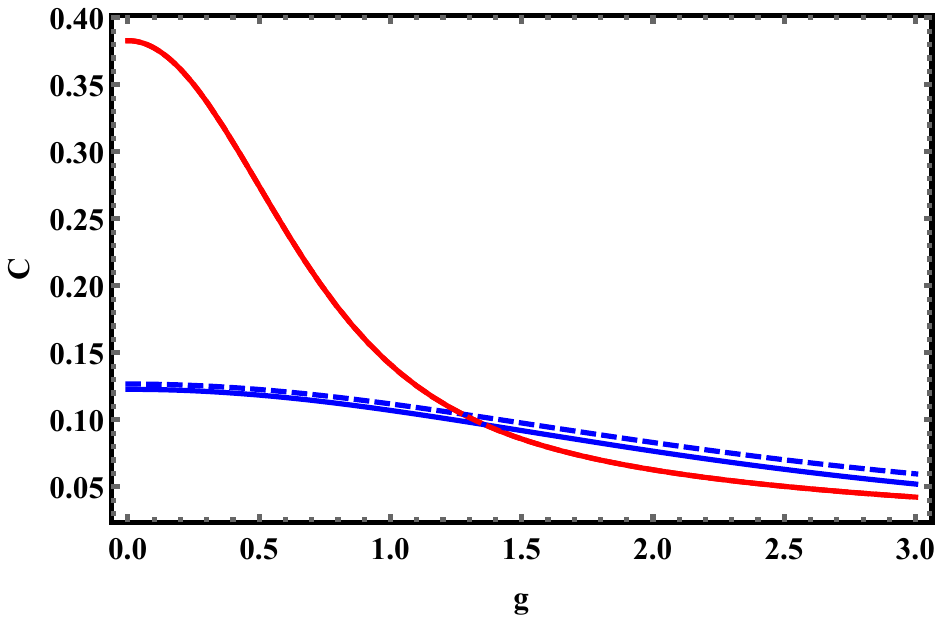}
 \includegraphics[width=0.45\linewidth]{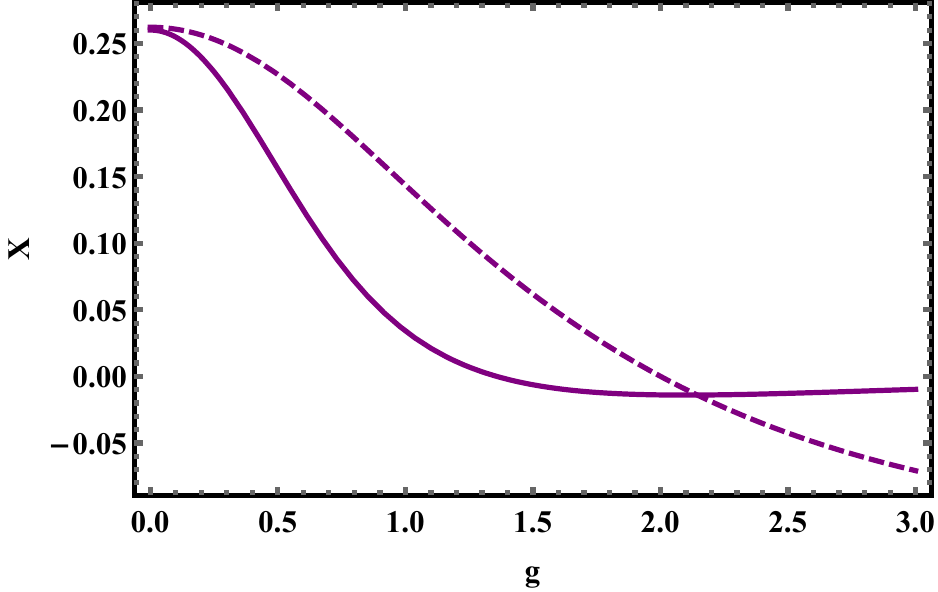}
 \caption{(i) Concurrence a s function of atom-field interaction for hot thermalization (red) and cold thermalization (blue) as a function of atom-field interaction (ii) Change in correlation (solid) and scaled work (dotted) as a function of atom-field interaction for $n=3, T_h = 4, T_c =1, k = 0.1, J = 1$. }
\label{C vs g} 
\end{figure}

To begin with, we focus on the Stirling engine, which consists of two isothermal and two isochoric strokes, as previously mentioned. Heat exchange occurs in each stage of this engine. In Figure (\ref{C vs g}), we depict the correlation variation as a function of atom-field coupling for the different stages of the Stirling engine.
It is evident that entanglement is at a minimum during stages 1 and 2, corresponding to hot thermalization, and increases during stages 3 and 4, representative of cold thermalization. The correlation aligns with cold thermalization overlaps, a trend also noticeable during hot thermalization. A portion of the reduction in coherence is employed to perform work in the heat engine. A close interrelation exists between work done and the change in coherence, as useful work extraction is impeded when there is a negative change in correlation.

Both correlation and work exhibit decreasing trends concerning atom-field interaction, as depicted in Figure (ii). This decline occurs because when atom-field coupling dominates over spin-spin interaction, the absorbed energy from the bath is directed towards the atom-field coupling, resulting in negative work.
\section{Conclusion}
In this study, we performed a comparative analysis of quantum Otto and Stirling engines using atom-photon interactions as the working substance. By investigating two-level and four-level atomic systems confined in an optical cavity, we examined how quantum properties such as coherence and entanglement influence thermodynamic performance. Our results indicate that the Stirling engine generally achieves higher efficiency and greater work output than the Otto engine, with entanglement playing a key role in energy exchange processes. Additionally, we found that atom-field coupling has a significant impact on work extraction, with optimal performance occurring within a specific coupling regime. These findings enhance the understanding of quantum heat engines and their operational principles, offering insights into quantum energy conversion mechanisms and their relevance to quantum thermodynamics


\begin{thebibliography}{99}
\bibitem{qhe1}H. Scovil, E. Schulz-DuBois, \emph{Phys. Rev. Lett.} 2, 262 (1959)
\bibitem{qhe2}S. Vinjanampathy, J. Anders, \emph{Contemp. Phys.} 57  545–579 (2016).
\bibitem{qhe_book} F. Binder, L.A. Correa, C. Gogolin, J. Anders, G. Adesso, Thermodynamics in the
Quantum Regime, vol. 195, Springer, 2018.
\bibitem{qhe3} M.T. Mitchison, \emph{Contemp. Phys.} 60 (2019) 164–187.
\bibitem{adv1} Y.H. Shi, H.L. Shi, X.H. Wang, M.L. Hu, S.Y. Liu, W.L. Yang, H. Fan, J. Phys. A,
Math. Theor. 53 (2020) 085301.
\bibitem{adv2} K. Brandner, M. Bauer, M.T. Schmid, U. Seifert, \emph{New J. Phys.} 17 (2015) 065006.
\bibitem{adv3} R. Uzdin, A. Levy, R. Kosloff, \emph{Phys. Rev. X} 5 (2015) 031044.
\bibitem{adv4} R. Uzdin, \emph{Phys. Rev. Appl.} 6 (2) (2016) 024004.
\bibitem{adv5}F. Altintas, A.Ü. Hardal, Ö.E. Müstecaplioglu, \emph{Phys. Rev. E} 90 (2014) 032102.
\bibitem{adv6}T. Zhang, W.T. Liu, P.X. Chen, C.Z. Li, \emph{Phys. Rev. A} 75 (2007) 062102.
\bibitem{adv7}F. Altintas, A.Ü. Hardal, Ö.E. Müstecaplioglu, \emph{Phys. Rev. A} 91 (2015) 023816.
\bibitem{adv8} G. Thomas, M. Banik, S. Ghosh, \emph{Entropy} 19 (2017) 442.
\bibitem{coh_bath1} D. Türkpenc¸ e, O.E. Müstecaplioglu, \emph{Phys. Rev. E} 93 (2016) 012145.
\bibitem{coh_bath2} D. Türkpenc¸ e, F. Altintas, M. Paternostro, Ö.E. Müstecaplioglu,  \emph{Europhys. Lett.}
117 (2017) 50002.
\bibitem{coh_bath3} M.O. Scully, M.S. Zubairy, G.S. Agarwal, H. Walther, \emph{Science} 299 (2003)
862–864.
\bibitem{coh_bath4} A.Ü. Hardal, Ö.E. Müstecaplioglu, \emph{Sci. Rep.} 5 (2015) 1–9.
\bibitem{coh_bath5} H. Quan, P. Zhang, C. Sun, \emph{Phys. Rev. E }73 (2006) 036122.
\bibitem{sq_bath1} X. Huang, T. Wang, X. Yi, et al., \emph{Phys. Rev. E} 86 (2012) 051105.
\bibitem{sq_bath2} J. Roßnagel, O. Abah, F. Schmidt-Kaler, K. Singer, E. Lutz, \emph{Phys. Rev. Lett}. 112
(2014) 030602.
\bibitem{sq_bath3} W. Niedenzu, V. Mukherjee, A. Ghosh, A.G. Kofman, G. Kurizki, \emph{Nat. Commun.}
9 (2018) 1–13.
\bibitem{non_mar_bath1} X. Zhang, X. Huang, X. Yi, \emph{J. Phys. A, Math. Theor.} 47 (2014) 455002.
\bibitem{corr_bath1} R. Dillenschneider, E. Lutz, \emph{Europhys. Lett.} 88 (2009) 50003.
\bibitem{Stirling1}Tien D. Kieu.,  \emph{J. Appl. Phys.},
90 (2001) 3086–3089.
\bibitem{Stirling2}G. S. Agarwal and S. Chaturvedi, emph{Phys. Rev. E},
88 (2013) 012130.
\bibitem{Stirling3}H. Xiao-Li et. al, \emph{Eur. Phys. J. D}, 68 (2014) 1434 .
\bibitem{JC_Model} E. T. Jaynes and F. W. Cummings, \textit{Proced.  IEEE}, 51  (1963) 89-109.
\bibitem{TC_model}M. Tavis and F. k W. Cummings, \emph{Phys. Rev.}, 170  (1968) 379–384.
\bibitem{Cocurrence} W. K Wotters, \emph{Phys. Rev. Lett.}, 80  (1997)  2245.
\end{thebibliography}
\end{document}